\documentclass[a4paper,11pt]{article}%
\usepackage{amsmath,url}%
\begin{document}

\title{The F.A.S.T.-Model
\author{Tobias Kretz and Michael Schreckenberg \\ \\
Physik von Transport und Verkehr\\ Universit\"{a}t Duisburg-Essen\\ D-47048 Duisburg, Germany
}
}

\maketitle       

\begin{abstract}
A discrete model of pedestrian motion is presented that is implemented in the \em Floor field- and Agentbased Simulation Tool (F.A.S.T.) \em which has already been applicated to a variety of real life scenarios.
\end{abstract}

\section{The F.A.S.T.-Model of Pedestrian Motion}
The F.A.S.T. model is discrete in space and time with an orthogonal lattice. It can be classified as probabilistic CA with extensions demanded by reality. There is a hard-core exlusion between the agents, of which at maximum one can stand at a cell at a certain point in time. An agent needs the space of one cell. This implies a cell size of roughly $40 \cdot 40$ $cm^2$, the minimum space a pedestrian occupies \cite{Dreyfuss67}. So far the model follows earlier models \cite{Kluepfel03,Kessel01}. In fact this model is in many aspects an extension - mainly related to speeds larger one cell per round - of the model presented in \cite{Kirchner02} which itself had predecessors \cite{Burstedde01a,Schadschneider01a,Burstedde01b}.\par
There are three levels of decision making in this model: 1) The choice of an exit, 2) the choice of a \em destination cell\em, 3) the path between the current and the destination cell. The first two are probabilistic processes. The third one is deterministic, ex\-cept for the order in which the agents carry out their steps to reach the des\-ti\-na\-tion cell. 
The process of choosing a destination cell is done completely in parallel by all agents, while the actual motion is a totally sequential process.
\par
In the following a {\em round} includes the decision for an exit as well as for a destination cell and all {\it steps}, while a {\it step} is the movement of an agent from one cell to one of the nearest neighbour cells i.e. a part of the path from the current towards the destination cell.
\subsection{Choosing an Exit} \label{definition}
At the beginning of each round all agents choose one of the exits with the probability $p_E^A=N(1+\delta_{AE}k_E(A))/S(A,E)^2$, with $A$ numbering the agents, $E$ numbering the exits agent $A$ is allowed to use, $\delta_{AE}=1$ if agent $A$ chose exit $E$ during the last round and $\delta_{AE}=0$ otherwise, $k_E(A)$ being agent $A$'s persistance to stick with a once taken decision for one of the exits, $S(A,E)$ being the distance between the exit and the current position of agent $A$, and $N$ as normalization constant guaranteeing $\sum_E p_E = 1$. The distance is squared so the probability is proportional to the inverse of the area of a circle around the exit with radius $S(A,E)$. Given a homogeneous density of agents all over a scenario with high symmetry this area is proportional to the num\-ber of agents which are closer to the exit than agent $A$. Therefore this is a measure of a possible queue before agent $A$ at exit $E$.\\
\subsection{Choosing a Destination Cell}
In a model which is spatially and temporally discrete an agent's (dimensionless) velocity is the number of cells which he is allowed to move during one round. As the real-world interpretation of the size of a cell is fixed by the scale of the discretization, the real-time interpretation of one round fixes the real-world interpretation of such a dimensionless velocity. One round is chosen to equal the typical reaction time of one secound. Typical maximal velocities $v_{max}$ of the agents therefore are three to six cells per round.\\
In the F.A.S.T. model an agent chooses one cell (the destination cell) he wants to move to out of all cells he would be able to reach during one round, except for those that are occupied. Which cells are part of the neighbourhood that belongs to a certain $v_{max}$ (i.e. the shape of such a neighbourhood) is described in \cite{Kretz05}.

\subsubsection{Probabilities for the possible Destination Cells:} 
Probabilities get assigned to each free and unoccupied cell in the neighbourhood of an agent that corresponds to the maximum velocity of that agent, that that particular cell is chosen as destination cell. The probability that an agent chooses cell $(x,y)$ is
\begin{equation}
p=Np_{xy}^Sp_{xy}^Dp_{xy}^Ip_{xy}^Wp_{xy}^P
\end{equation}
While $N$ is a normalization constant all $p_{xy}^X$ are partial probabilities from the different influences on the movement of an agent.
\begin{enumerate}
\item $p_{xy}^S$ is the influence of the \em static floor field \em which contains the information on the distance towards the exit.
\item $p_{xy}^D$ is the influence of the \em dynamic floor field \em \cite{Schadschneider01b} which contains the information of the motion of the other agents.
\item $p_{xy}^I$ is the influence of inertia effects.
\item $p_{xy}^W$ is the influence of nearby walls.
\item $p_{xy}^P$ is the influence of the density of nearby agents.
\end{enumerate}
These five influences will be introduced in more detail now.
\paragraph{Moving towards the Exit - Following the Static Floor Field:}
Before the simulation begins, the distance from each cell to each exit is calculated using Dijkstra's algorithm \cite{Dijkstra59} and stored in the static floor field. With the static floor field $p^S$ is calculated for a certain cell at $(x,y)$ as $p^S_{xy}=e^{-k_SS_{xy}}$, with $k_S$ being the \em coupling strength \em of an agent to the static floor field knowledge as well as will to move are parametrized. All of the five influences are weighted against each other in their relative strengths by \em coupling constants \em $k_X$ and all coupling constants are individual parameters of the agents.
\paragraph{Herding Behaviour - Following Others:}
Asides the main CA where the agents move, there is another CA - \em the dynamic floor field \em - where agents leave a virtual trace whenever they move. This trace decays and diffuses with time. In the F.A.S.T. model the dynamic floor field is a vectorial field. So an agent who has moved from $(a,b)$ to $(x,y)$ changes the dynamic floor field $(D_x,D_y)$ at $(a,b)$ by $(x-a,y-b)$ after all agents have moved. Right after that all values of both components of $D$ decay with probability $\delta$ and diffuse with probability $\alpha$ to one of the (von Neumann) neighbouring cells. Since the vector components can be negative, decay means a reduction of the absolute value. Diffusion is only possible from x- to x- and from y- to y-component. The influence on the motion of the agents is $p^D_{xy}=e^{k_D(D_x(x,y)(x-a)+D_y(x,y)(y-b))}$ where $(a,b)$ is the current position of the agent.
\paragraph{Inertia:} Contrary to Newtonian physics pedestrians experience de- and accelerating in motion direction as being less ardous than walking through curves. Due to the shape and functionality of the human movement apparatus pedestrians can de- and accelerate from and to their preferred walking velocities almost  instantaneously compared to a timescale of one secound. However deviating quickly by e.g. $90^\circ $ from a certain direction while keeping up the walking speed is far more difficult. So only centrifugal forces are considered to have an influence on the motion of the agents.\\
On a perfect circle the centrifugal force - which in the F.A.S.T. model is the measure for inertia influences - is $F_c \propto v^2/r$. This assumption after a few steps \cite{Kretz06} leads to the following inertia dependence of the movement probability $p^I(x_{t+1},y_{t+1}) = e^{- k_I (v_{t+1}+v_{t}) \sin{\frac{|\phi|}{2}}}$ with $\phi$ as angle of deviation from the former direction of motion and $t$ counting the timesteps.
\paragraph{Safety Distance towards Walls.} This is considered via $p^W_{xy}=e^{(-k_WW_{xy})}$ where $W_{xy}$ is the distance of the cell $(x,y)$ towards the closest wall. For distances larger than a certain $W_{max}$ the effect vanishes completely and $p^W_{xy}=1$.\\
\paragraph{Staying Polite - Keeping a Distance Towards other Agents.}
After each round for each cell $(x,y)$ the number $N_P(x,y)$ of agents in its Moore neighbourhood is counted. The more agents are immediately neighboured, the less another agent might want to choose this cell as his destination: $p^P_{xy}=e^{-k_PN_P(x,y)}$.
Then the density at the border of a crowd changes less rapidly, while in the center of the crowd the density remains high since all free cells which an agent can reach during one round are surrounded by agents.

\subsection{Moving towards the Destination Cell}
Once all agents have chosen their destination cell the agents start moving towards them. The sequence in which the agents execute their steps is chosen randomly in a way, that agent D executes a step, then agent B, then again agent D, then agent K. During one step an agent moves deterministically onto that cell within the Moore neighbourhood of his current cell, that lies closest to his destination cell. To represent the dynamic space consumption \cite{Weidmann93} a cell which has been occupied once during this process remains blocked for all other agents until the end of the round. So if the cell closest to the destination cell is blocked the agent moves to the secound closest and so on. If there is no cell left that is unblocked and closer to the destination cell than the current cell the round ends for that particular agent.

\end{document}